\documentclass[twocolumn,prb,superscriptaddress,noshowpacs,epsf]{revtex4}

\usepackage{graphicx}

\begin{document}
\title{
Controllable manipulation and entanglement of macroscopic quantum states
in coupled charge qubits}

%
%
\author{J. Q. You}
\affiliation{Frontier Research System, The Institute of Physical
and Chemical Research (RIKEN), Wako-shi 351-0198, Japan}
\author{J. S. Tsai}
\affiliation{Frontier Research System, The Institute of Physical and
Chemical Research (RIKEN), Wako-shi 351-0198, Japan}
\affiliation{NEC Fundamental Research Laboratories, Tsukuba, Ibaraki 305-8051,
Japan}
\altaffiliation[Permanent address.]{}
\author{Franco Nori}
\altaffiliation[Corresponding author~(Email:~nori@umich.edu).]{}
\affiliation{Frontier Research System, The Institute of Physical and
Chemical Research (RIKEN), Wako-shi 351-0198, Japan}
\affiliation{Center for Theoretical Physics, Physics Department, Center
for the Study of Complex Systems, The University of Michigan, Ann Arbor,
MI 48109-1120, USA}
\altaffiliation[Permanent address.]{}

\begin{abstract}
We present an experimentally implementable method to couple
Josephson charge qubits and to generate and detect macroscopic
entangled states. A large-junction superconducting quantum interference 
device is used in the qubit circuit for
both coupling qubits and implementing the readout.
Also, we explicitly show how to achieve a microwave-assisted macroscopic 
entanglement in the coupled-qubit system.
\end{abstract}
\maketitle

\section{Introduction}

Quantum-mechanical systems can exploit the fundamental properties of
superposition and entanglement to process information in an efficient and
powerful way that no classical device can do. Recently, Josephson-junction
circuits have received renewed attention because these
may be used as qubits in a quantum
computer.~\cite{review} Based on the charge and phase degrees of freedom
in Josephson-junction devices, charge~\cite{NPT99,MSS} and
phase qubits~\cite{MOOIJ,VAL,FRIED} have been developed. Also,
a type of solid-state qubit can be realized in a large-area
current-biased Josephson junction.~\cite{MART,HAN} 

Experimentally, coherent
oscillations were demonstrated in a Josephson charge qubit prepared
in a superposition of two charge states.~\cite{NPT99}
More recent experimental measurements~\cite{VION}
showed that the charge qubit at suitable working points
can have a sufficiently high quality of coherence
($Q_{\varphi}\approx 2.5\times 10^4$),
corresponding to a decoherence time $T_{\varphi}\approx 500$~ns.
Current-biased Josephson junctions can also have long decoherence 
times~\cite{HAN,MART} and $Q_{\varphi}$ can reach $10^4$.
These exciting experimental advancements
demonstrate the potential of Josephson qubits for manufacturing
macroscopic quantum-mechanical machines. Towards the practical implementation
of a solid-state quantum computer, the next important step would be the
coupling of two qubits and then scaling up the
architecture to many qubits.

In this work, we present an experimentally implementable method to couple
two Josephson charge qubits and to generate and detect macroscopic
quantum entangled states in this charge-qubit system.
Motivated by very recent experimental results,~\cite{VION}
we employ a superconducting quantum interference device (SQUID) with {\it two}
large Josephson junctions to implement the readout.
The generation of the macroscopic entanglement
is assisted by applying a microwave field to each charge qubit.
The key advantage of our design is that
the SQUID can also produce an experimentally feasible and controllable
coupling between the two charge qubits. As verified in a
single qubit,~\cite{VION} the coupled charge qubits may be well decoupled
from the readout system when the measurement is not implemented.
Moreover, our design can be readily extended
to coupled multiple~\cite{YTN} qubits as well as any selected pairs
(not necessarily neighbors).

The paper is organized as follows. In Sec.~II, the controllable
coupling between two charge qubits is proposed using a large Josephson 
junction or a large-junction dc SQUID. Also, we demonstrate how this interbit
coupling can be conveniently used to generate the controlled-phase-shift gate. 
In Sec.~III, we study the microwave-assisted macroscopic quantum 
entanglement in the coupled charge qubits, where the microwave fields are 
coupled to the qubits via gate capacitances. Section IV focuses on the readout
of the quantum states in the coupled-qubit system. Finally, the discussion and 
conclusion are given in Sec.~V.

\subsection{Other qubit coupling schemes}

A different type of interbit coupling from the one studied here 
was proposed using the Coulomb
interaction between charges on the islands of the charge 
qubits.~\cite{PFP} As pointed out in Ref.~\onlinecite{review}, 
the interbit coupling in this scheme is not switchable
and also it is hard to make the system scalable because only neighboring
qubits can be coupled.
Implementations of quantum algorithms such as the Deutsch and
Bernstein-Vazirani algorithms were studied using a system of Josephson
charge qubits,~\cite{SIE} where it was proposed that the nearest-neighbor
superconducting islands would be coupled by tunable dc SQUIDs.
In Ref.~\onlinecite{BLAIS}, a pair of charge qubits were proposed to be 
capacitively coupled to a current-biased Josephson junction 
where, by varying the bias current, the junction can be tuned in and out 
of resonance with the qubits coupled to it.

Another different type of interbit coupling was proposed~\cite{review,MSS}
in terms of the oscillator modes in an $LC$ circuit. In contrast, we use 
a large junction or a large-junction dc SQUID (but {\it no} $LC$ circuit) 
to couple the charge qubits. In our scheme, 
{\it both} dc and ac supercurrents can flow through the charge-qubit circuit, 
while in Refs.~\onlinecite{review} and \onlinecite{MSS} 
{\it only} ac supercurrents can flow through 
the circuit. These yield different interbit couplings (e.g., the
$\sigma_y\sigma_y$ type~\cite{review,MSS} as opposed to $\sigma_x\sigma_x$
in our proposal). As revealed in 
Ref.~\onlinecite{YTN}, the $\sigma_x\sigma_x$-type 
interbit coupling can be conveniently used to formulate an efficient 
quantum-computing scheme. 

Moreover, the calculated interbit-coupling terms 
in Refs.~\onlinecite{review} and 
\onlinecite{MSS} only apply to the case in which the following two conditions are 
met: 

(i) The eigenfrequency $\omega_{LC}$ of the $LC$ circuit is much faster 
than the quantum manipulation frequencies. This condition {\it limits} the 
allowed number $N$ of the qubits in the circuit because $\omega_{LC}$ scales 
with $1/\sqrt{N}$. In other words, this implies that the circuits in 
Refs.~\onlinecite{review} and \onlinecite{MSS} are not really scalable.

(ii) The phase conjugate to the total charge
on the qubit capacitors fluctuates weakly. Our interbit-coupling 
approach discussed below is free from these two limitations.

\begin{figure}
\includegraphics[width=3.2in,bbllx=106,bblly=374,bburx=568,bbury=651]
{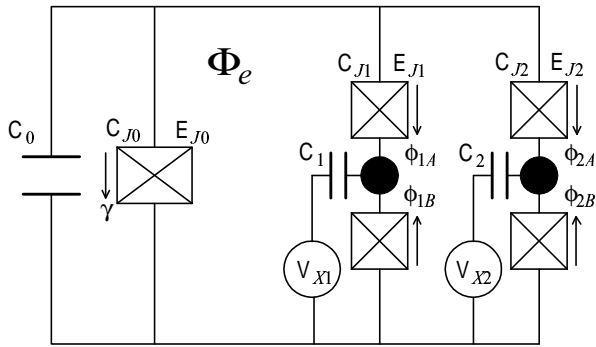}
\caption{Schematic diagram of two charge qubits coupled by a large 
Josephson junction (denoted by a square with a ``$\times$" inside) 
of coupling energy $E_{J0}$ and capacitance $C_{J0}$.
To make the effective charging energy of the large Josephson junction 
as small as required, a large capacitance $C_0$ is placed close to 
and in parallel with it.
Each filled circle denotes a superconducting island, the Cooper-pair box,
which is biased by a voltage $V_{Xi}$ via the gate capacitance
$C_i$ and coupled to the bulk superconductors by two identical small 
Josephson junctions (each with a coupling energy $E_{Ji}$ and a capacitance 
$C_{Ji}$). Here the arrow near each Josephson junction denotes the chosen 
direction for the positive phase drop across the corresponding junction.}
\label{F1}
\end{figure}

\section{Controllable coupling of charge qubits}

\subsection{Coupling qubits with a large junction}

We first use a large Josephson junction to couple
two charge qubits (see Fig.~\ref{F1}). Each qubit is realized
by a Cooper-pair box, where a superconducting island with excess
charge ${\hat Q}_i=2e{\hat n}_i$ ($i=1,2$) is weakly coupled to
the bulk superconductors via two identical small junctions (with Josephson
coupling energy $E_{Ji}$ and capacitance $C_{Ji}$) and biased by
an applied voltage $V_{Xi}$ through a gate capacitance $C_i$. The large
Josephson junction on the left 
has a coupling energy $E_{J0}$ (much larger than $E_{Ji}$)
and a capacitance $C_{J0}$. 
As in the single-qubit case,\cite{VION} close to the large Josephson 
junction, we also place a large capacitance $C_0$ in parallel with it, so that 
the effective charging energy of the large Josephson junction 
can be ignored (even though the capacitance of the large 
junction might not be large enough). Moreover, we assume that the
inductance of the qubit circuit (i.e., the two Cooper-pair boxes with
the nearby junctions, and the superconducting lines connecting these two qubits 
with the large Josephson junction) is much
smaller than the Josephson inductance of the large junction. 
The Hamiltonian of the system  can be written as
\begin{eqnarray}
&&H=\sum_{i=1}^2\left[E_{ci}(\hat{n}_i-n_{Xi})^2-E_{Ji}(\cos\hat{\phi}_{iA}
+\cos\hat{\phi}_{iB})\right] \nonumber\\
&&~~~~~~~~~~~ -E_{J0}\cos\hat{\gamma},
\label{AA}
\end{eqnarray}
where 
\begin{equation}
E_{ci}={2e^2\over C_i+2C_{Ji}}
\end{equation} 
is the charging energy of the superconducting island and 
$n_{Xi}=C_iV_{Xi}/2e$ is the reduced offset charge
(in units of $2e$) induced by the gate voltage. 
Flux quantization around loops containing the phase drops of the involved 
junctions gives the constraint:
\begin{equation}
\hat{\phi}_{iA}-\hat{\phi}_{iB}-\hat{\gamma}+{2\pi\Phi_e\over \Phi_0}=0,
~~~~ i=1,2,
\end{equation}
which gives
\begin{eqnarray}
\hat{\phi}_{iA}\!&\!=\!&\!\hat{\phi}_i-
\left({\pi\Phi_e\over\Phi_0}-{1\over 2}\hat{\gamma}\right),\nonumber\\
\hat{\phi}_{iB}\!&\!=\!&\!\hat{\phi}_i+
\left({\pi\Phi_e\over\Phi_0}-{1\over 2}\hat{\gamma}\right),
\end{eqnarray}
where the average phase drop 
$\hat{\phi}_i={1\over 2}(\hat{\phi}_{iA}+\hat{\phi}_{iB})$
is canonically conjugate to the number, $\hat{n}_i$, of the excess Cooper 
pairs on the $i$th superconducting island:
\[
[\hat{\phi}_j,\hat{n}_j]=i, \:\:\: j=1,2.
\]
Here $\hat{\phi}_{iA}$ and $\hat{\phi}_{iB}$ ($i=1,2$) are the phase drops 
across the small Josephson junctions above (A) and below (B) the $i$th 
Cooper-pair box.  

The Hamiltonian (\ref{AA}) can be 
rewritten as
\begin{eqnarray}
&&H=\sum_{i=1}^2\left[E_{ci}(\hat{n}_i-n_{Xi})^2 
-2E_{Ji}\cos\left({\pi\Phi_e\over\Phi_0}
-{1\over 2}\hat{\gamma}\right)\right.\nonumber\\
&&~~~~~~~~~~~~~~~~~~~\left.\times\cos\hat{\phi}\right] 
-E_{J0}\cos\hat{\gamma}.
\label{BB}
\end{eqnarray}  
The externally applied flux $\Phi_e$ threads the area between the large 
Josephson junction and the left Cooper-pair box. 
It induces circulating supercurrents 
in the qubit circuit. The total circulating supercurrent $\hat{I}$ has
contributions from the two charge qubits: 
\begin{equation}
\hat{I}=\hat{I}_1+\hat{I}_2, 
\end{equation}
where
\begin{equation}
\hat{I}_i=2I_{ci}\sin\left({\pi\Phi_e\over\Phi_0}
-{1\over 2}\hat{\gamma}\right)\cos\hat{\phi}_i,
\end{equation}
with $I_{ci}=\pi E_{Ji}/\Phi_0$. This total supercurrent flows through the
large Josephson junction and it can also be written as 
\begin{equation}
\hat{I}=I_0\sin\hat{\gamma},
\end{equation}
with $I_0=2\pi E_{J0}/\Phi_0$. From Eqs.~(6)-(8) it follows that
\begin{equation}
I_0\sin\hat{\gamma}=
2\sin\left({\pi\Phi_e\over\Phi_0}-{1\over 2}\hat{\gamma}\right)
(I_{c1}\cos\hat{\phi}_1+I_{c2}\cos\hat{\phi}_2).
\label{A1}
\end{equation}
When the coupling energy $E_{Ji}=\Phi_0I_{ci}/\pi$ 
of each Josephson junction connected to the 
charge box is much smaller than that of the large Josephson junction
in the circuit, the phase drop $\hat{\gamma}$ across the large junction 
will be small. Expanding the operator functions of $\hat{\gamma}$ 
in Eq.~(\ref{A1}) into a series and retaining the terms up to second order
of the parameters 
\begin{equation}
\eta_i={I_{ci}\over I_0}\:(<1), \:\:\: i=1,2, 
\end{equation}
we have
\begin{eqnarray}
\hat{\gamma}\!&\!=\!&\!2\sin\left({\pi\Phi_e\over\Phi_0}\right)
\left(\eta_1\cos\hat{\phi}_1 +\eta_2\cos\hat{\phi}_2\right) \nonumber\\
&&\!-\sin\left({2\pi\Phi_e\over\Phi_0}\right)
\left(\eta_1\cos\hat{\phi}_1+\eta_2\cos\hat{\phi}_2\right)^2.
\end{eqnarray} 
It is clear that the phase drop $\hat{\gamma}$ across the large Josephson
junction is controllable via the applied flux $\Phi_e$.

For Hamiltonian (\ref{BB}), we also expand the operator functions of 
$\hat{\gamma}$ into a series and retain the terms up to second order
of $\eta_i$. Moreover, we consider the charging regime with $E_{ci}$ much
larger than $E_{Ji}$. 
Also, we assume that the temperature is low enough ($k_BT\ll E_{ci}$)
and the superconducting gap is larger than $E_{ci}$, so that quasiparticle
tunneling is strongly suppressed. In this case, only the lowest two charge
states are important for each qubit operating around the degeneracy point
$V_{Xi}=(2n_i+1)e/C_i$. 
In the spin-$1\over 2$ representation based on the charge states
$|n_i\rangle\equiv|\!\uparrow\rangle_i$, and
$|n_i+1\rangle\equiv|\!\downarrow\rangle_i$ of each Cooper-pair box,
the Hamiltonian of the system can be reduced to
\begin{equation}
H=\sum_{i=1}^2\left[\varepsilon_i(V_{Xi})\,\sigma_z^{(i)}
-\overline{E}_{Ji}\,\sigma_x^{(i)}\right]
-\chi\,\sigma^{(1)}_x\sigma^{(2)}_x,
\label{CC}
\end{equation}
with
\begin{equation}
\varepsilon_i(V_{Xi})={1\over 2}E_{ci}\left[{C_iV_{Xi}\over e}-(2n_i+1)\right].
\end{equation}
and
\begin{equation}
\overline{E}_{Ji}=E_{Ji}\cos\left({\pi\Phi_e\over\Phi_0}\right)\xi_i,
\end{equation}
where
\begin{equation}
\xi_i=1-{3\over 8}(\eta_i^2+3\eta_j^2)
\sin^2\left({\pi\Phi_e\over\Phi_0}\right),
\end{equation}
and $i,j=1,2$ ($i\ne j$). 
The interbit coupling $\chi$ is given by
\begin{equation}
\chi=L_JI_{c1}I_{c2}\sin^2\left({\pi\Phi_e\over\Phi_0}\right),
\end{equation} 
where the large Josephson junction acts as an {\it effective} inductance of 
value
\begin{equation}
L_J={\Phi_0\over 2\pi I_0}.
\end{equation}
It is clear that the interbit coupling is switched off at $\Phi_e=0$. 
It is well known that a large Josephson junction can act as an inductance
(e.g., Ref.~\onlinecite{review}). 
Here we explicitly show a specific way that it can be used to couple qubits.

Retaining up to second-order terms in the expansion parameters $\eta_i$,
the total circulating current $\hat{I}$ can be written as
\begin{eqnarray}
\hat{I}\!&\!=\!&\!2\sin\left({\pi\Phi_e\over\Phi_0}\right)
(I_{c1}\cos\hat{\phi}_1+I_{c2}\cos\hat{\phi}_2) \nonumber\\
&-\!&\!{1\over I_0}\sin\left({2\pi\Phi_e\over\Phi_0}\right)
(I_{c1}\cos\hat{\phi}_1+I_{c2}\cos\hat{\phi}_2)^2.
\end{eqnarray}
In the spin-${1\over 2}$ representation, it is given by
\begin{eqnarray}
\hat{I}\!&\!=\!&\!\sin\left({\pi\Phi_e\over\Phi_0}\right)
(I_{c1}\sigma^{(1)} _x+I_{c2}\sigma^{(2)}_x) \nonumber\\
&-\!&\!{1\over 4I_0}\sin\left({2\pi\Phi_e\over\Phi_0}\right)
\left[I_{c1}^2+I_{c2}^2+2I_{c1}I_{c2}\sigma^{(1)}_x\sigma^{(2)}_x\right],\nonumber\\
&&
\label{DD}
\end{eqnarray}
which depends on the states of the charge-qubit system.

\begin{figure}
\includegraphics[width=3.4in,bbllx=15,bblly=360,bburx=568,bbury=680]
{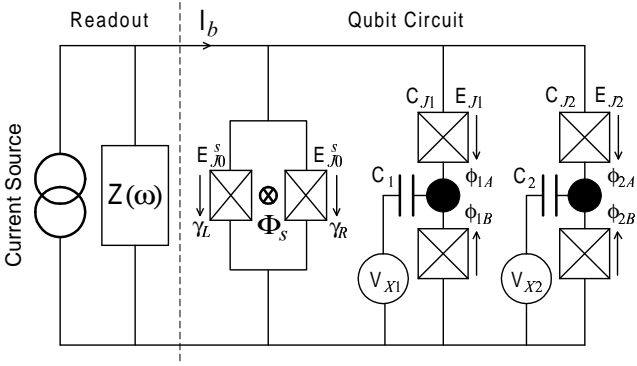}
\caption{Schematic diagram of the coupled-qubit circuit with
a biased-current source of impedance $Z(\omega)$. The
dc SQUID, with two junctions of large $E_{J0}^s$, plays the role of 
both coupling the charge qubits and implementing the readout. 
Here the large capacitance $C_0$ placed
close to and in parallel with the dc SQUID is included in the impedance
$Z(\omega)$.}
\label{F2}
\end{figure}

\subsection{Coupling qubits with a SQUID}

There are somewhat conflicting requirements imposed on this circuit.
To obtain a large value for the effective Josephson inductance 
$L_J=\Phi_0/2\pi I_0$, 
a relatively small $I_0$ is needed, so that a large interbit coupling can be 
achieved. 
However, when the large Josephson junction is also employed for a 
readout, it is desirable to use a large $I_0$.
This permits a larger range of $I_b$, so that a higher resolution in
distinguishing qubit states can be achieved in the quantum measurement 
based on the switching of the supercurrent through the large junction. 

These two opposite requirements can be conveniently solved if the leftmost 
large Josephson junction in Fig.~\ref{F1} is 
replaced by a symmetric dc SQUID with two sufficiently large 
junctions (see Fig.~\ref{F2}). Instead of $\Phi_e$ inside the circuit loop
between $E_{J0}$ and the first qubit (as in Fig.~1), 
we now apply a flux $\Phi_s$ inside the large-junction 
dc SQUID loop (see Fig.~2). 
This SQUID can be used both for coupling the two charge qubits and 
implementing the readout.
When the readout is not active ($I_b=0$), we can choose a suitable flux 
$\Phi_s$ inside the SQUID loop to generate a larger interbit coupling. 
For $I_b=0$, the reduced Hamiltonian of the
coupled-qubit system and the total circulating current $I$ have the same 
forms as in Eqs.~(\ref{CC}) and (\ref{DD}), 
but with $\Phi_e$ and $I_0$ replaced 
by ${1\over 2}\Phi_s$ and 
\begin{equation}
I_0=2I_0^s\cos\left({\pi\Phi_s\over\Phi_0}\right),
\end{equation} 
where $I_0^s=2\pi E_{J0}^s/\Phi_0$.   
When the readout is active (see Sec.~IV), $\Phi_s$ is chosen as zero to obtain
a larger effective Josephson coupling energy.

\subsection{Controlled-phase-shift gate}

When the system works at the degeneracy points with
$\varepsilon_i(V_{Xi})=0$, the Hamiltonian becomes
\begin{equation}
H=-\overline{E}_{J1}\,\sigma_x^{(1)}-\overline{E}_{J2}\,\sigma_x^{(2)}
-\chi\,\sigma^{(1)}_x\sigma^{(2)}_x.
\end{equation}
For instance, when $\overline{E}_{Ji}>0$, $i=1,2$, its four eigenvalues
are 
\begin{eqnarray}
&&\overline{E}_{J1}+\overline{E}_{J2}-\chi,\nonumber\\
&&\overline{E}_{J1}-\overline{E}_{J2}+\chi,\nonumber\\
&&\overline{E}_{J2}-\overline{E}_{J1}+\chi,\nonumber\\
&&-\overline{E}_{J1}-\overline{E}_{J2}-\chi.
\end{eqnarray}
The corresponding eigenstates are $|e_1,e_2\rangle$,
$|e_1,g_2\rangle$, $|g_1,e_2\rangle$,
and $|g_1,g_2\rangle$, where
\begin{eqnarray}
|e_i\rangle\!&\!=\!&\!{1\over\sqrt2}
(|\!\uparrow\rangle_i-|\!\downarrow\rangle_i),
\nonumber\\
|g_i\rangle\!&\!=\!&\!{1\over\sqrt2}
(|\!\uparrow\rangle_i+|\!\downarrow\rangle_i).
\end{eqnarray}
Because they are also the eigenstates of the two uncoupled
charge qubits, when prepared initially at
an eigenstate, the system does not evolve to an entangled state
even in the presence of interbit coupling.
As shown below, one can take advantage of this property to
implement the measurement. In addition, this property can be
used to construct efficient conditional gates. For instance,
if 
\begin{equation}
\overline{E}_{J1}=\overline{E}_{J2}=\chi,
\end{equation}
the controlled-phase-shift (CPS) gate is given by
\begin{equation}
U_{\rm CPS}(\tau)=e^{i\chi\tau/\hbar}U, 
\end{equation}
with
\begin{eqnarray}
U\!&\!=\!&\!e^{-iH\tau/\hbar} \nonumber\\
&\!=\!&\!\exp\left.\{i(\chi\tau/\hbar)[\sigma^{(1)}_x
+\sigma^{(2)}_x+\sigma^{(1)}_x\sigma^{(2)}_x]\right.\},
\end{eqnarray}
at $\tau=\pi\hbar/4\chi$. This gate transforms the basis states 
$|e_1,e_2\rangle$, $|e_1,g_2\rangle$, $|g_1,e_2\rangle$, 
and $|g_1,g_2\rangle$ as
\begin{equation}\left(
\begin{array}{c}
|e_1,e_2\rangle\\ 
|e_1,g_2\rangle\\ 
|g_1,e_2\rangle\\ 
|g_1,g_2\rangle
\end{array}\right)
\longrightarrow\left(
\begin{array}{cccc}
1 & 0 & 0 & 0 \\
0 & 1 & 0 & 0 \\
0 & 0 & 1 & 0 \\
0 & 0 & 0 & -1 
\end{array}\right)
\left(
\begin{array}{c}
|e_1,e_2\rangle\\ 
|e_1,g_2\rangle\\ 
|g_1,e_2\rangle\\ 
|g_1,g_2\rangle
\end{array}\right).
\end{equation}
The generation of this conditional two-bit gate is efficient 
because the condition (24) can be realized in one step via changing the gate
voltages $V_{Xi}$, $i=1,2$, and the flux $\Phi_s$ simultaneously. Also,
the architecture is scalable because multiple charge qubits can be coupled
by connecting them in parallel with the large-junction SQUID. 
If the two Josephson junctions
in each Cooper-pair box are replaced by small-junction dc SQUIDs,
any selected pairs of charge qubits (not necessarily neighbors) can be 
coupled.~\cite{YTN}

\section{Microwave-assisted macroscopic entanglement} 

When a
microwave field is applied to the Josephson charge qubit, Rabi oscillations
occur in the system.~\cite{NPT02} These oscillations can also be
demonstrated by coupling a quantum resonator to the charge
qubit.~\cite{ABS} Here we apply the microwave field to the Cooper-pair box
via the gate capacitance, as in Refs.~\onlinecite{VION} and \onlinecite{NPT02}, 
but each charge qubit is driven by a different microwave field.~\cite{YN}
In this situation, $n_{Xi}$ in Eq.~(\ref{AA})
is replaced by
\begin{equation}
n_{Xi}+\hat{n}_{ACi}=n_{Xi}+\left({C_id_i\over 2e}\right)\hat{{\cal E}}_{ACi}.
\end{equation}
Here $d_i$ is the thickness of the gate capacitor and 
\begin{equation}
\hat{{\cal E}}_{ACi}=
{\cal E}_{\lambda i}a_i+{\cal E}^*_{\lambda i}a_i^{\dag}
\end{equation}
is the microwave electric field in the gate capacitor of the $i$th
Cooper-pair box,
where $a_i$ is the annihilation operator of the microwave mode. 
Because the microwave wavelength is much larger than
$d_i$, ${\cal E}_{\lambda i}$ can be considered constant in the gate
capacitor. In the charging regime,
the Hamiltonian of the system (including the microwave fields) can be
written as
\begin{eqnarray}
&&H=\sum_{i=1}^2\left[\varepsilon_i(V_{Xi})\,\sigma_z^{(i)}
-\overline{E}_{Ji}\,\sigma_x^{(i)}+\hbar\omega_{\lambda i}
\,a_ia_i^{\dag}\right.\nonumber\\
&&~~~~~~~~~~~\left.+\sigma_z^{(i)}\,(K_i a_i+K^*_i a_i^{\dag})\right]
-\chi\,\sigma^{(1)}_x\sigma^{(2)}_x,
\end{eqnarray}
where 
\begin{equation}
K_i=\left(E_{ci}C_id_i\over 2e\right){\cal E}_{\lambda i}. 
\end{equation}
Here, we also consider the
system working at the degeneracy points $\varepsilon_i(V_{Xi})=0$, $i=1,2$.
When $\hbar\omega_{\lambda i}\approx 2|\overline{E}_{Ji}|$
and under the rotating-wave approximation, the Hamiltonian is cast to 
\begin{eqnarray}
H\!&\!=\!&\!\sum_{i=1}^2\left[
-\overline{E}_{Ji}\,\sigma_x^{(i)}+\hbar\omega_{\lambda i}
\,a_ia_i^{\dag}+(K_i|e_i\rangle\langle g_i|a_i+{\rm H.c.})\right] \nonumber\\
&&~~~~~~-\chi\,\sigma^{(1)}_x\sigma^{(2)}_x.
\end{eqnarray}
Without interbit coupling, each Josephson charge qubit exhibits
Rabi oscillations between states $|e_i,l_i\rangle$ and $|g_i,l_i\!+\!1\rangle$,
where $|l_i\rangle$ is a photon state with $l_i$ photons. For the resonant
case with $\hbar\omega_{\lambda i}=2|\overline{E}_{Ji}|$, the eigenvalues of
each charge-qubit system are given by
\begin{equation}
\epsilon^{(i)}_{\pm}=E_{0i}\pm{1\over 2}\hbar\Omega_i,
\end{equation}
where 
\begin{equation}
E_{0i}=\hbar\omega_{\lambda i}(l_i+1), 
\end{equation}
and
\begin{equation}
\Omega_i={2\over\hbar}|K_i|\sqrt{l_i+1} 
\end{equation}
is the Rabi frequency.
Though entanglement occurs between each charge qubit and the nonclassical
microwave field, the two qubits do not entangle with each other
since the system evolves as
\begin{equation}
|\Psi(t)\rangle=|\psi_1(t)\rangle|\psi_2(t)\rangle,
\end{equation}
where 
\begin{equation}
|\psi_i(t)\rangle=\sin(\Omega_it)|e_i,l_i\rangle
+\cos(\Omega_it)|g_i,l_i\!+\!1\rangle
\end{equation}
if the system is initially prepared at state
$|g_1,g_2,l_1\!+\!1,l_2\!+\!1\rangle$.
However, in the presence of microwave fields,
when the interbit coupling is switched on, the coupled-qubit
system exhibits complicated quantum oscillations and it will evolve to the
entangled state.
For instance, in the resonant situation, the eigenvalues are given by
\begin{eqnarray}
\varepsilon_{1,4}\!&\!=\!&\!E_{01}+E_{02}\pm\hbar\Lambda_1, \nonumber\\
\varepsilon_{2,3}\!&\!=\!&\!E_{01}+E_{02}\pm\hbar\Lambda_2,
\end{eqnarray}
where 
\begin{equation}
\Lambda_{1,2}=[(\Omega_1\pm\Omega_2)^2+(\chi/\hbar)^2]^{1/2}.
\end{equation}

The state of the coupled-qubit system evolves as
\begin{eqnarray}
|\Psi(t)\rangle\!&\!=\!&\!C_1(t)|e_1,e_2,l_1,l_2\rangle
+C_2(t)|e_1,g_2,l_1,l_2\!+\!1\rangle \nonumber\\
&&\!+C_3(t)|g_1,e_2,l_1\!+\!1,l_2\rangle \nonumber\\
&&\!+C_4(t)|g_1,g_2,l_1\!+\!1,l_2\!+\!1\rangle.
\label{EE}
\end{eqnarray}
For the system prepared initially at
$|g_1,g_2,l_1\!+\!1,l_2\!+\!1\rangle$,
\begin{eqnarray}
C_1(t)\!&\!=\!&\!{1\over 2}\{R_2(t)-R_1(t)
+i(\chi/\hbar)[S_2(t)-S_1(t)]\},\nonumber\\
C_2(t)\!&\!=\!&\!{1\over 2}[(\Omega_1+\Omega_2)S_1(t)
+(\Omega_1-\Omega_2)S_2(t)],\nonumber\\
C_3(t)\!&\!=\!&\!{1\over 2}[(\Omega_1+\Omega_2)S_1(t)
-(\Omega_1-\Omega_2)S_2t)],\nonumber\\
C_4(t)\!&\!=\!&\!{1\over 2}\{R_1(t)+R_2(t)
+i(\chi/\hbar)[S_1(t)+S_2(t)]\}, \nonumber\\
&&
\end{eqnarray}
where 
\begin{equation}
R_i(t)=\cos(\Lambda_it), ~~~~ S_i(t)={\sin(\Lambda_it)\over\Lambda_i}.
\end{equation}
For a two-level system interacting with a single-mode field, the Rabi
oscillations can be explained using either quantum or semiclassical theory,
where the single-mode field is described quantum mechanically or treated as
a classical field.~\cite{SCULLY} 
Here the quantum oscillations of coupled charge qubits 
(namely, the Rabi oscillations in coupled two-level systems) 
are studied using quantum theory, where the microwave field coupled to 
each qubit is quantized. This also applies to the classical-field case, 
in which the quantum oscillations are still described by
Eq.~(\ref{EE}), but $|e_1,e_2,l_1,l_2\rangle$, $|e_1,g_2,l_1,l_2+1\rangle$,
$|g_1,e_2,l_1+1,l_2\rangle$, and $|g_1,g_2,l_1+1,l_2+1\rangle$
are replaced by $|e_1,e_2\rangle$, $|e_1,g_2\rangle$, $|g_1,e_2\rangle$,
and $|g_1,g_2\rangle$.

\begin{figure}
\includegraphics[width=3.3in,bbllx=75,bblly=280,bburx=512,bbury=762]
{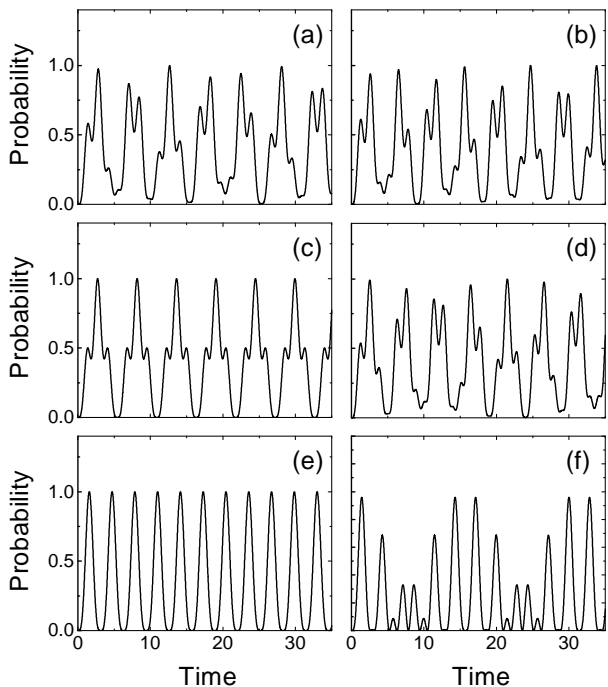}
\caption{Occupation probability $|C_1(t)|^2$ as a function of time.
(a)~$\Omega_2=\Omega_1,\chi/\hbar=\Omega_1$; (b)~$\Omega_2=1.2\Omega_1,
\chi/\hbar=\Omega_1$; (c)~$\Omega_2=\Omega_1,
\chi/\hbar=\sqrt{3}\Omega_1/2$; (d)~$\Omega_2=1.2\Omega_1,
\chi/\hbar=\sqrt{3}\Omega_1/2$; (e)~$\Omega_2=\Omega_1,\chi=0$;
(f)~$\Omega_2=1.2\Omega_1,\chi=0$.
The time is in units of $\Omega_1^{-1}$.}
\label{F3}
\end{figure}

Figure~\ref{F3} shows the occupation probability $|C_1(t)|^2$ as a function
of time $t$. For instance, when $|C_1(t)|^2\simeq 1$, both charge qubits are 
in their excited states. 
It can be seen that $|C_1(t)|^2$ looks very different when
the interbit coupling is switched on or off.
The macroscopic entanglement between the two coupled qubits can be
explicitly shown at $\Omega_1=\Omega_2$ $(=\Omega)$. In this case,
when $t_{\rm ent}=n\pi\hbar/W\chi$, with $n=1,2,3, \dots$, and
\begin{equation}
W=[(2\hbar\Omega/\chi)^2+1]^{1/2},
\end{equation}
$|\Psi(t)\rangle$ becomes 
\begin{eqnarray}
|\Psi(t_{\rm ent})\rangle\!&\!=\!&\!C_1(t_{\rm ent})|e_1,e_2,l_1,l_2\rangle \nonumber\\
&&\!+C_4(t_{\rm ent})|g_1,g_2,l_1+1,l_2+1\rangle, 
\end{eqnarray}
where
\begin{eqnarray}
C_1(t_{\rm ent})\!&\!=\!&\!{1\over 2}[-\cos(n\pi)+\exp(in\pi/W)], \nonumber\\
C_4(t_{\rm ent})\!&\!=\!&\!{1\over 2}[\cos(n\pi)+\exp(in\pi/W)].
\end{eqnarray}
The peaks away from either zero or 1 
shown in Fig.~\ref{F3}(a) correspond to this kind of
entangled state. Furthermore, if suitable values of $W$ are taken,
the maximally entangled state with $|C_1|^2=|C_4|^2={1\over 2}$
can be derived. This state is a macroscopic
Schr{\"o}dinger-cat state of the two charge qubits.
For instance, if $\hbar\Omega/\chi=\sqrt{3}/2$, the coupled-qubit 
system evolves to the maximally entangled state
at the times given by 
\begin{equation}
t_{\rm ent}^{\rm(max)}=(2l+1)\pi\hbar/2\chi,~~~~ l=0,1,2,\dots. 
\end{equation}
This entangled
state corresponds to the half-probability peaks in Fig.~\ref{F3}(c).

\section{Quantum measurement}

To implement a readout,
we bias a current pulse $I_b$ to the qubit circuit (see Fig.~\ref{F2}),
as in the single-qubit case.~\cite{VION} Now, a term
$-\Phi_0I_b\hat{\delta}/2\pi$,
with 
\begin{equation}
\hat{\delta}={1\over 4}[\hat{\gamma}_L+\hat{\gamma}_R
+\sum_{i=1,2}(\hat{\phi}_{iA}-\hat{\phi}_{iB})],
\end{equation}
should be added to the Hamiltonian (\ref{AA}),
where $\hat{\delta}$ is the average phase drop of the total qubit circuit
and it can be written as 
\begin{equation}
\hat{\delta}=\hat{\gamma}-{\pi\Phi_s\over 2\Phi_0},
\end{equation}
with 
$\hat{\gamma}={1\over 2}(\hat{\gamma}_L+\hat{\gamma}_R)$.
Here we set the flux $\Phi_s$ equal to zero
to have a larger effective Josephson coupling energy.
In the spin-${1\over 2}$ representation based on charge states,
the Hamiltonian of the system is also reduced to Eq.~(\ref{CC}). The interbit
coupling is here induced by the bias current and given by
\begin{equation}
\chi=L_JI_{c1}I_{c2}\sin^2(\gamma_0/2),
\end{equation}
where the effective inductance is
\begin{equation}
L_J={\Phi_0\over 2\pi I_0\cos\gamma_0}, 
\end{equation}
and 
\begin{equation}
\gamma_0=\sin^{-1}({I_b/I_0}),
\end{equation}
with $I_0=4\pi E_{J0}^s/\Phi_0$, and $I_b<I_0$.
The intrabit couplings are
\begin{equation}
\overline{E}_{Ji}=E_{Ji}\cos(\gamma_0/2)\xi_i,
\end{equation}
where 
\begin{equation}
\xi_i=1-\alpha(\eta_i^2+3\eta_j^2)\sin^2(\gamma_0/2),
\end{equation}
with 
\begin{equation}
\alpha={{2+\cos\gamma_0}\over 8\cos^3\gamma_0}, 
\end{equation}
and $i,j=1,2$ $(i\ne j)$.
The supercurrent through the SQUID,
\begin{eqnarray}
I_0\sin\hat{\gamma}\!&\!=\!&\!I_b-\sin(\gamma_0/2)
[I_{c1}\sigma_X^{(1)}+I_{c2}\sigma_X^{(2)}] \nonumber\\
&+\!&\!{1\over 4I_0}\tan\gamma_0[I_{c1}^2+I_{c2}^2
+2I_{c1}I_{c2}\sigma_X^{(1)}\sigma_X^{(2)}], \nonumber\\
&&
\end{eqnarray}
has contributions from both the bias current
and the current from the Josephson charge qubits.

\begin{figure}
\includegraphics[width=3.2in,bbllx=80,bblly=355,bburx=520,bbury=720]
{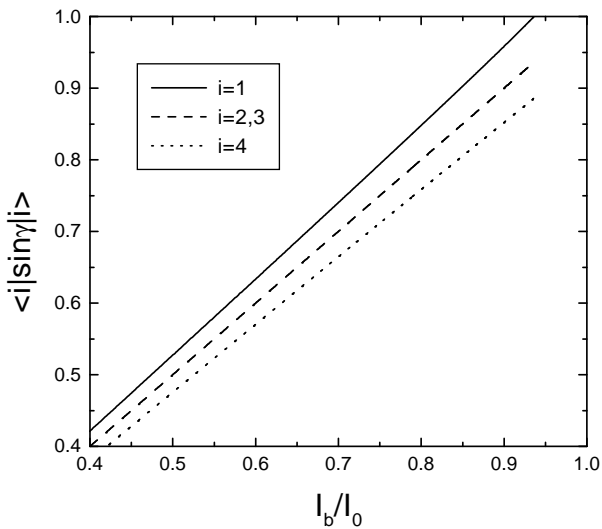}
\caption{Eigenstate dependence of the supercurrent through the SQUID
as a function of the bias current $I_b$.
Here, $E_{J1}=E_{J2}={1\over 5}E_{J0}^s$, $|1\rangle=|e_1,e_2\rangle$,
$|2\rangle=|e_1,g_2\rangle$, $|3\rangle=|g_1,e_2\rangle$,
and $|4\rangle=|g_1,g_2\rangle$.}
\label{F4}
\end{figure}

At the working points with $\varepsilon_i(V_{Xi})=0$, 
the eigenstates of the system are also
$|e_1,e_2\rangle$, $|e_1,g_2\rangle$,
$|g_1,e_2\rangle$, and $|g_1,g_2\rangle$.
In Fig.~\ref{F4}, we show the dependence of the supercurrents through 
the SQUID on the eigenstates of the charge-qubit system. The supercurrents
through the SQUID increase with 
the bias current
and the difference between the supercurrents at different (nondegenerate)
eigenstates widens.
For the measurement setup shown in Fig.~\ref{F2}, the supercurrent through
the SQUID is the largest at the eigenstate $|e_1,e_2\rangle$
and it first reaches the maximal value $I_0$ (namely, the critical
current) when the bias current $I_b$ approaches a value $I_{SW}$
near $I_0$. Around this value, the supercurrent through the SQUID
switches, with a very large probability $P_1$,
from the zero-voltage state to the dissipative nonzero-voltage state in the
quasiparticle-current branch and the measurement on the voltage is
carried out. However, due to environmental noise as well as
thermal and quantum fluctuations,
the switching actually occurs before the supercurrent through
the SQUID reaches $I_0$.
At $I_b\sim I_{SW}$, the supercurrents through the SQUID will also
switch to the nonzero voltage state at other eigenstates, but the switching
probabilities are small. In the ideal case,
if the difference between the large switching
probability $P_1$ and the small ones is close to 1, then, in principle, 
a single-shot readout would be achieveable.
As shown in Ref.~\onlinecite{VION}, the Josephson-junction switching
experiment can provide sufficient accuracy to discriminate
the state $|e_1,e_2\rangle$ from others.

The operation and readout of the macroscopic entanglement of the
coupled-qubit system can be implemented by simultaneously applying
a pulsed microwave field (with the same duration $\tau$) to each charge
qubit. The sequence would be: 

(i)~before the microwave fields are applied,
the flux $\Phi_s$ through the SQUID is set equal to zero and no interbit
coupling exists; 

(ii)~the flux $\Phi_s$ is switched on to
a certain nonzero value
exactly at the start of the microwave pulse and off at the end of
the microwave pulse. Within the microwave pulse duration $\tau$,
the evolution of the system is described by Eq.~(\ref{EE}); 

(iii)~a pulsed
bias current $I_b$ is applied to perform a measurement after
the microwave pulse. 

During the measurement, the quantum state of
the charge-qubit system collapses to the eigenstate
$|e_1,e_2\rangle$ with probability $|C_1(\tau)|^2$. This
probability is proportional to the switching probability $P_1$ of
the SQUID. Because of relaxation, the envelope of the measured
switching probability $P_1$
decays exponentially with time. This is used to obtain the relaxation
time.\cite{NPT99,VION} Ramsey fringes of the probability
$P_1$ can be used~\cite{VION} to determine the decoherence time
of the coupled-qubit system.
For each given microwave pulse duration
$\tau$, through repeated measurements, one can determine the
occupation probability $|C_1(\tau)|^2$ and thus deduce the
information about the macroscopic entanglement between the coupled
charge qubits [see Figs.~3(a) and 3(c)].

\section{Discussion and Conclusion}

Finally, we estimate some important parameters using available
quantities for the single charge qubit. Here we consider
the maximally entangled case shown in Fig.~\ref{F3}(c), in which
$\Omega_1=\Omega_2=\Omega$, and 
\[
{\chi\over\hbar}={\sqrt{3}\over 2}\Omega.
\]
Taking $2\pi/\Omega\simeq 0.22$~$\mu$s, as derived from the
Rabi oscillation of the measured switching probability,~\cite{VION}
we have $\chi/\hbar\approx 0.25$~GHz. Reference~\onlinecite{VION} also gives
$E_J/\hbar\approx 16.5$~GHz.
Choosing 
\[
E_{J0}^s\approx 5E_{Ji}\approx 5E_J,
\] 
and using the
relation for $\chi$, we obtain $\Phi_s\approx 0.35\Phi_0$.
For $\Phi_s=0$, the expansion parameters are
\[
\eta_i={I_{0i}\over I_0}\approx 0.05
\] 
for $E_{J0}^s\approx 5E_{Ji}$.
When $\Phi_s\approx 0.35\Phi_0$, they become $\eta_i\approx 0.14$.
The results are sufficiently accurate when $\overline{E}_{Ji}$
and $\chi$ are retained up to second- and higher-order terms in the
expansion parameters $\eta_i$. When $\Phi_s$ approaches $\Phi_0/2$,
the interbit coupling strengthens. The reduced Hamiltonian of
the system also has the same form as Eq.~(\ref{CC}), but higher-order terms
in the expansion parameters should be included to obtain accurate
results.

Here we consider the charging regime 
with $E_{ci}\gg E_{Ji}$ in order to obtain analytical results. We expect that 
the interbit coupling can still be realized in the regime with 
$E_{ci}\sim E_{Ji}$, i.e., the regime used by the Saclay group in the 
experiment on a single Josephson qubit.~\cite{VION} In this latter regime, 
the results can only be obtained numerically, but a relatively
long decoherence time would be expected for the coupled-qubit system 
to work at the degeneracy points because at these points 
the states are more stable against the variations of both the offset charges 
and the flux $\Phi_e$ or $\Phi_s$.  

Very recently, quantum oscillations were experimentally observed in two coupled 
charge qubits.~\cite{PASH} Also, a novel method for the controllable coupling 
of charge qubits was proposed using a variable electrostatic transformer.~\cite{Averin}
In contrast with our interbit coupling scheme, these studies involve 
capacitively-coupled (as opposed to inductively-coupled) charge qubits. The main 
advantage of this inductive coupling among qubits is that it allows a controllable 
link between any selected qubits, not necessarily nearest neighbors.  

In conclusion, we employ a large-junction SQUID to couple
Josephson charge qubits and implement a readout.
This architecture is readily scalable to multiple qubits.
When the system works at the degeneracy
points, where the dephasing effects are suppressed, it is shown that
the macroscopic entanglement can be generated with the assistance of
microwave fields.
Also, we show the quantum measurement of the macroscopic entanglement.

\begin{acknowledgments}
We thank X. Hu and B. Plourde for useful comments and acknowledge support from
the U.S. ARDA, AFOSR, and the U.S. National Science Foundation
grant No.~EIA-0130383.
\end{acknowledgments}

\end{document}